\documentclass[twocolumn,
aps,showpacs,preprintnumbers,amsmath,amssymb,floatfix,prl]{revtex4}

\usepackage{color,amsmath,graphicx,pst-text,%
amssymb,
            pst-grad,pstricks,pst-3d
}
\usepackage{leftidx}
\input epsf
\usepackage{epsfig}
\begin{document}

\newcommand{\alphaFn}[1]{\alpha\left(#1\right)}
\newcommand{\Laplace}[2]{\mathcal{L}\left\lbrace #1\right\rbrace\left(#2\right)}
\newcommand{\Mellin}[2]{\mathcal{M}\left\lbrace #1\right\rbrace\left(#2\right)}
\newcommand{\Fourier}[2]{\mathcal{F}\left[ #1\right]\left(#2\right)}
\newcommand{\FourierC}[2]{\mathcal{F}_{c}\left[ #1\right]\left(#2\right)}
\newcommand{\Fourierb}[2]{\mathcal{F}\left\lbrace #1\right\rbrace\left(#2\right)}
\newcommand{\FourierCb}[2]{\mathcal{F}_{c}\left\lbrace #1\right\rbrace\left(#2\right)}
\newcommand{\Fracderiv}[1]{\mathcal{D}^{1-#1}}
\newcommand{\FracIntN}[1]{\frac{\partial^{#1}}{\partial t^{#1}}}
\newcommand{\FracIntM}[2]{\frac{\partial^{#1} #2 }{\partial t^{#1}}}
\newcommand{\FracderivN}[1]{\frac{\partial^{1-#1}}{\partial t^{1-#1}}}
\newcommand{\FracderivM}[2]{\frac{\partial^{1-#1} #2 }{\partial t^{1-#1}}}
\newcommand{\RLderivO}[1]{\frac{d^{#1}}{dt^{#1}}}
\newcommand{\RLderivOM}[2]{\frac{d^{#1} #2}{dt^{#1}}}
\newcommand{\del}{\partial}

\title{Reaction-subdiffusion on moving fluids.}

\author{H. Zhang}
\email{zhanghong13@cdut.cn}
\affiliation{Department of Mathematics Teaching, Chengdu University of
Technology, Cheng du 610059, China.}

\author{G.H. Li}
\email{liguohua13@cdut.cn}
\affiliation{Department of Mathematics Teaching, Chengdu University of
Technology, Cheng du 610059, China.}

\date{\today}
\begin{abstract}
To capture the dynamic behaviors of reaction-subdiffusion in flow
fields, in the present paper we analyze a simple monomolecular
conversion $A\rightarrow B$. We derive the corresponding master
equations for the distribution of A and B particles in continuous time random walks scheme. The new results are then
 used to obtain the generalizations of
 advection-diffusion
reaction equation, in which the diffusion and advection operators both depend on the reaction rate.
\end{abstract}
\maketitle

Reactive transport in flows is an
important issue of diffusion theory that has a variety of
applications in many topics such as the transport of contaminants in
underground water \cite{SKB2002}, nuclear waste storage \cite{ABB2015}, etc. The macroscopic description of reaction-diffusion in a
velocity field is the standard advection-diffusion
reaction equation (ADRE) defined in one-dimensional form as:
\begin {equation}
\frac{\partial C(x,t)}{\partial t}+v\frac{\partial C(x,t)}{\partial
x}=K\frac{\partial^2 C(x,t)}{\partial x^2}+f
\end {equation}
where $C(x,t)$ is the  probability density function (PDF) of the
particle, $v$ is constant velocity, $K$ is diffusion coefficient, and $f$ denotes the decoupled reaction term.

In recent years the reaction under anomalous diffusion has attracted
more and more attention \cite{SSS2006}. One of the effective ways to capture
anomalous diffusion is continuous time random walk (CTRW) model
\cite{MK2000,MK2004,BMB2011,ZL2016}. By using CTRW Sokolov et al. analyzed reaction-subdiffusion
schemes for the monomolecular conversion A $\rightarrow$ B, derived
the corresponding kinetic equations for local A and B concentrations, and first argued that reaction-subdiffusion equations are not
obtained by a trivial change of the diffusion operator for a
subdiffusion one \cite{SSS2006}. Such nontrivial coupled effect was also found
in some other reaction-subdiffusion systems \cite{FS2008,CM2009,AYL2010,ADH2013}, for instance, on
the form of stationary solutions for reaction-diffusion in finite
domains (see Ref.~\cite{FS2008}).

However, up to now few works have approached the
reaction under anomalous diffusion in nonhomogeneous flows \cite{C2015,SPB2017,BBS2017}.
 Without considering the effect of chemical reactions, Compte \cite{C1997,CMC1997} has discussed
anomalous diffusion in a nonhomogeneous
convection velocity field  by applying the
CTRW techniques in which the step length distribution function
  depends on the starting point of the jump, and showed
   that convection coefficient depends on the waiting time statistics.
But to our knowledge,  the more basic and interesting problem of how
nonhomogeneous velocity field might affect the reaction under
anomalous diffusion is still unknown. This is what we shall address in this
paper. In what follows we will consider the reaction-subdiffusion
process on moving fluid  for the reaction A$\rightarrow$B in CTRW
scheme, then derive the generalizations of advection-diffusion
reaction equation and show the interesting coupling relations
between the velocity field and the chemical reactions under
subdiffusion.

We start by recalling the CTRW model on inhomogeneous flows in
one-dimensional case \cite{C1997,CMC1997}. In this model, the jump length $y$ for
the moving particle is dragged along the velocity $v(x)$ and
replaced by $ y-\tau_{a}v(x)$
 where $\tau_{a}$ stands for an advection
time scale, and $\tau_{a}v(x)$ is the mean drag experienced by a
particle jumping from the point $x$. Thus, the particle jumps from
$x$ to $x+y$ with the jump length PDF $\lambda(y-\tau_{a}v(x))$, and
then waits at $x+y$ for time $t$ drawn from $\psi(t)$, after which
the process is renewed.

We then consider the simplest reaction scheme $A\rightarrow B$ in
this CTRW model. We assume all properties of A and B particles are
the same and the particles trapped in stagnant regions will react
with a relabeling of A into B taking place at a rate $\alpha$. Let
$A(x,t)$ be the PDF of A particle being in point $x$ at time $t$ and
and $i^{-}(x,t)$ be the escape rate. By assuming that in the initial
distribution all particles have zero resting times, we can find the
balance equation for A particles in a given point:
\begin{eqnarray}\label{A(x,t)}
A(x,t)&=&A_{0}(x)\Psi(t)e^{-\alpha
t}+\int_{-\infty}^{+\infty}dx'\int_{0}^{t}i^{-}(x',t')\nonumber \\
                    & &\lambda(x-x'-\tau_{a}v(x'))\Psi(t-t')e^{-\alpha(t-t')}dt'
\end{eqnarray}
where $A_0(x)$ is  the initial state  of A particle,
$\Psi(t)e^{-\alpha t}=(1-\int_{0}^{t}\psi(\tau)d\tau)e^{-\alpha t}$
is the joint survival density of remaining at least at time $t$ on
the spot (without being converted into B). The density is a sum of
outgoing particles from all other points at different times given by
the flow, and provided they survived after their arrival till the
time $t$. The first term on the right hand side is just the
influence of the initial distribution.

The above
equation (2) can be changed to the form
\begin{eqnarray}\label{A(x,t)e}
A(x,t)&=&A_{0}(x)\Psi(t)e^{-\alpha
t}+\int_{-\infty}^{+\infty}dx'\int_{0}^{t}i^{-}(x',t')\nonumber\\
& &\phi(x-x',t-t';x')e^{-\alpha(t-t')}dt'
\end{eqnarray}
by using the expression $\phi(r,\tau;x)=\lambda(r-\tau_{a}v(x))\Psi(\tau)$ \cite{C1997}.
Fourier transforming $x\rightarrow k$ and Laplace transforming
$t\rightarrow u$ of Eq.(3), we obtain
\begin{equation}\label{A(k,u)}
A(k,u)=A_{0}(k)\Psi(u+\alpha)+\int
i^{-}(k',u)\phi(k,u+\alpha;k-k')dk'.
\end{equation}
Here, $A_0(k)$ represents the Fourier
$x\rightarrow k$ transform  of the initial condition $A_0(x)$,
$\Psi(u+\alpha)$ denote the Laplace transform of joint survival PDF
$\Psi(t)e^{-\alpha t}$, $i^{-}(k,u)$ is the Fourier-Laplace
transform of $i^{-}(x,t)$,
 and
\begin{eqnarray}
\phi(k,u+\alpha;k-k')&=&\Psi(u+\alpha)\lambda(k)\nonumber\\
& &\int
e^{-ik\tau_{a}v(x')}e^{-i(k-k')x'}dx'.
\end{eqnarray}
To obtain the master equation with respect to $A(x,t)$, we shall give the other balance equation.
Noticing
that the loss flux is from those particles that were originally at
$x$ at $t =0$ and wait without reacting until time $t$ to leave, and
those particles that arrived at an earlier time $t'$ and wait
without reacting until time $t$ to leave, we have the second
balance equation:
\begin{eqnarray}\label{i-(x,t)}
&i^{-}(x,t)=A_{0}(x)\psi(t)e^{-\alpha
t}+\int_{-\infty}^{+\infty}dx'\int_{0}^{t}i^{-}(x',t')\nonumber\\&~~\times
\lambda(x-x'-\tau_{a}v(x'))\psi(t-t')e^{-\alpha(t-t')}dt'
\end{eqnarray}
where $\psi(t)e^{-\alpha t}$ is the non-proper waiting time density
for the actually made new step provided the particle survived \cite{SSS2006}.
 By introducing
$$\eta(r,\tau;x)=\lambda(r-\tau_{a}v(x))\psi(\tau)$$ and
applying the transform $(x,t)\rightarrow (k,u)$ of Eq.(6), we find
\begin{eqnarray}\label{i-(k,u)}
i^{-}(k,u)&=&A_{0}(k)\psi(u+\alpha)\nonumber\\
& &+\int
i^{-}(k',u)\eta(k,u+\alpha;k-k')dk'
\end{eqnarray}
where the term $\eta(k,u+\alpha;k-k')=\psi(u+\alpha)\lambda(k)\int
e^{-ik\tau_{a}v(x')}e^{-i(k-k')x'}dx'$. We divide  Equation (4) by
(7) to write
\begin{equation}
i^{-}(k,u)=\frac{\psi(u+\alpha)}{\Psi(u+\alpha)}A(k,u)
\end{equation}
Nothing that $\Psi(u+\alpha)=\frac{1-\psi(u+\alpha)}{u+\alpha}$, we
get
\begin{equation}
i^{-}(k,u)=\Phi_{\alpha}(u+\alpha)A(k,u)
\end{equation}
where
$\Phi_{\alpha}(u+\alpha)=\frac{(u+\alpha)\psi(u+\alpha)}{1-\psi(u+\alpha)}$,
which recovers the relation between $A(x,t)$ and $i^{-}(x,t)$ when
the effect of the flow field  is not considered in Ref.\cite{SSS2006}.
Inverting Eq.(9) to the space-time domain $k\rightarrow x$,
$s\rightarrow t$, we obtain
\begin{equation}
i^{-}(x,t)=\int_{0}^{t}\Phi_{\alpha}(t-t')A(x,t')dt'.
\end{equation}
Here, the kernel $\Phi_{\alpha}(t)$ is equal in laplace
$t\rightarrow u$ space to $\Phi_{\alpha}(u+\alpha)$.  When
$\alpha=0$, it reduces to  the usual memory kernel of master equation for CTRW \cite{SSS2006,HLS2010,FF2012}.

We now consider a linear velocity field $v(x)=\omega x$ where
$\omega$ is a constant. Then Eq.(4) becomes
\begin{equation}
A(k,u)=\Psi(u+\alpha)A_{0}(k)+\Psi(u+\alpha)\lambda(k)j(k+v_{k},u)
\end{equation}
where the symbol $v_{k}=\tau_{a}\omega k$.
 In the limit $\tau_{a}\rightarrow 0$, Eq.(11) gives
 \begin{eqnarray}
A(k,u)&\simeq&\Psi(u+\alpha)A_{0}(k)+\Psi(u+\alpha)\lambda(k)\nonumber\\
& &\times(i^{-}(k,u)+v_{k}i^{-}{'}_{k}(k,u)).
\end{eqnarray}
We substitute (9) into (12)  and get
\begin{eqnarray}
A(k,u)=\Psi(u+\alpha)A_{0}(k)+\psi(u+\alpha)\lambda(k)\nonumber\\
\times(A(k,u)+v_{k}A'_{k}(k,u)).
\end{eqnarray}
 This
simplifies further to the generalized master equation in
Fourier-Laplace space for A particles in an $A\rightarrow B$
reaction  under subdiffusion on linear moving fluid
\begin{eqnarray}
(1-\psi(u+\alpha)\lambda(k))A(k,u)=\Psi(u+\alpha)A_{0}(k)\nonumber\\
+\psi(u+\alpha)\lambda(k)v_{k}A'_{k}(k,u).
\end{eqnarray}
Notice that if the reaction  is not involved in the system and the
initial condition is defined
 as $A_{0}(x)=\delta(x)$, then Eq.(14) recovers the master equation
\begin{equation}
[1-\psi(u)\lambda(k)]A(k,u)=v_{k}\psi(u)\lambda(k)A'_{k}(k,u)+\Psi(u)
\end{equation}
  for the CTRW
 on linear moving fluids in one-dimensional lattice obtained by Compte in
 Ref.~\cite{C1997}.

There is the other way to derive the generalized master equation
(14) where the balance condition (3) is replaced by \cite{SSS2006}
\begin{equation}
\frac{\partial A(x,t)}{\partial t}=i^{+}(x,t)-i^{-}(x,t)-\alpha
A(x,t).
\end{equation}
Here, $i^{+}(x,t)$ is the gain flux which can be represented by the
loss flux \cite{FF2012}
\begin{equation}
i^{+}(x,t)=
 \int_{-\infty}^{+\infty}i^{-}(x',t)\lambda(x-x'-\tau_{a}v(x'))dx'.
 \end{equation}
Transforming $(x,t)\rightarrow(k,u)$ of (16), one has
\begin{eqnarray}
uA(k,u)-A_{0}(k)&=&\frac{\int
i^{-}(k',u)\eta(k,u+\alpha;k-k')dk'}{\psi(u+\alpha)}\nonumber\\
& &-i^{-}(k,u)-\alpha A(k,u)
\end{eqnarray}
By using (7) and (18), we can also obtain the relation equation (9).
Substitute Eq.(9) into Eq.(18) and assume $v(x)=\omega x$, in the
limit $\tau_{a}\rightarrow 0$, and we find
\begin{eqnarray}
uA(k,u)-A_{0}(k)&\simeq &(\lambda(k)\Phi_{\alpha}(u)-\Phi_{\alpha}(u))A(k,u)
+\lambda(k)\nonumber\\
& &\times\Phi_{\alpha}(u)v_{k}A'_{k}(k,u)-\alpha A(k,u)
\end{eqnarray}
Using
$\Phi_{\alpha}(u+\alpha)=\frac{(u+\alpha)\psi(u+\alpha)}{1-\psi(u+\alpha)}$
and $\Psi(u+\alpha)=\frac{1-\psi(u+\alpha)}{u+\alpha}$, one
finally recovers the generalized master equation (14). This means
that the two approaches to derive the generalized master equation
for A-particles in reaction-subdiffusion  process on moving fluid
are equivalent. In what follows we will not distinguish them for
A-particles.

Inverting Eq.(19) to the space-time domain, we can obtain the master equation in space-time
domain:
\begin{widetext}
\begin{eqnarray}
\frac{\partial A(x,t)}{\partial
t}+\tau_{a}\int_{-\infty}^{+\infty}dx'\int_{0}^{t}\Phi_{\alpha}(t-t')\lambda(x-x')\frac{\partial
v(x')A(x',t')}{\partial x'}dt'
&=&\int_{-\infty}^{+\infty}dx'\int_{0}^{t}\Phi_{\alpha}(t-t')[\lambda(x-x')\nonumber\\
& &-\delta(x-x')]A(x',t')dt'-\alpha
A(x,t),
\end{eqnarray}
\end{widetext}
 where the reaction rate explicitly affects both the transport term
 and the advection term.
Specifically, we consider a discrete random walk, where the PDF of
particles A on site $x = i$  at time $t$ is denoted as $A(i,t)$,
 and  the jump PDF is assumed to be $\lambda(-1)=\frac{1}{2},\lambda(1)=\frac{1}{2}$, meaning
 that the particle can jump from $x=i$ to the adjacent grid point, to the right and left
directions  with the same  probability. If $v(x)=0$, then Eq.(20)
reduces to
\begin{eqnarray}
\frac{\partial A(i,t)}{\partial t}&=&\int_{0}^{t}\Phi_{\alpha}(t-t')
[\frac{1}{2}A(i-1,t)\nonumber\\
& &+\frac{1}{2}A(i+1,t)-A(i,t)]dt'-\alpha A(i,t),
\end{eqnarray}
obtained by Sokolov in Ref.~\cite{SSS2006}.

We now turn to apply the master equation (19) to derive a
 ADRE for Gaussian jump length  $ \lambda(k)\sim
1-\frac{\sigma^{2}k^{2}}{2}$ and
 long-tailed  waiting time
$ \psi(u)\sim 1-\Gamma(1-\beta)(\tau u)^{\beta}$ with  $\tau$ and
$\sigma^{2}$ being the appropriate time scale and
  the jump length variance, respectively.
Assuming $\rho_{0}(x)=\delta(x)$, substituting
$\Phi_{\alpha}(u)=\frac{1}{\tau^{\beta}\Gamma(1-\beta)}(u+\alpha)^{1-\beta}$
\cite{SSS2006} into the master equation (19), in the limit of
$\tau\rightarrow 0, \sigma\rightarrow 0$ and $\tau_{a}\rightarrow 0$,
we obtain
\begin{eqnarray}
&uA(k,u)&-A_{0}(k)\simeq-K_{\beta}k^{2}(u+\alpha)^{1-\beta}A(k,u)
+\nonumber\\
& &C_{\beta}(u+\alpha)^{1-\beta}v_{k} A'_{k}(k,u)-\alpha A(k,u)
\end{eqnarray}
where the symbols $K_{\beta}=\lim_{\tau\rightarrow 0,
\sigma\rightarrow 0}\frac{\sigma^{2}}{2\tau_{\beta}\Gamma(1-\beta)}$
and $C_{\beta} =\lim_{\tau_{a}\rightarrow 0, \tau\rightarrow
0}\frac{\tau_{a}}{\tau_{\beta}\Gamma(1-\beta)}$ are kept finite.
 Inverting
Eq.(22) to the space-time domain $k\rightarrow x$ and $s\rightarrow
t$, using the fact that
$\mathcal{F}(xf(x))=i\frac{\partial}{\partial k}f(k)$, we then get
the generalized ADRE for A particle in reaction-subdiffusion process
on linear flows:
\begin{eqnarray}
\frac{\partial A(x,t)}{\partial
t}&&+C_{\beta}\hat{T}_{t}(1-\beta,\alpha)\frac{\partial(v(x)A(x,t))}{\partial
x}\nonumber\\
& &=K_{\beta}\hat{T}_{t}(1-\beta,\alpha)\frac{\partial^{2}A(x,t)}{\partial
x^{2}}-\alpha A(x,t)
\end{eqnarray}
with the initial condition $A_{0}(x)=\delta(x)$. Here, The integral
operator $\hat{T}_{t}(1-\beta,\alpha)f = \tau^{\beta}\Gamma
(1-\beta)\int_{0}^{t}\Phi_(\alpha)(t-t')f(t')dt' $ corresponds in
time domain to
\begin{eqnarray}
\hat{T}_{t}(1-\beta,\alpha)f&=&\frac{d
}{dt}\int_{0}^{t}\frac{e^{-\alpha(t-t')}}{(t-t')^{1-\beta}}f(t')dt'\nonumber\\
& &+\alpha\int_{0}^{t}\frac{e^{-\alpha(t-t')}}{(t-t')^{1-\beta}}f(t')dt'
\end{eqnarray}
and becomes a fractional derivative when $\alpha=0$ \cite{SSS2006}. It should
be noted that in the generalized ADRE (23) not only diffusion  but
also advection term depend on the reaction parameter $\alpha$.

Analogously we shall now derive the generalized ADRE for the
B-particles. Let $B(x,t)$ be the PDF of B particle being in point
$x$ at time $t$, $j^{+}(t)$ be the gain flux and $j^{-}(t)$ be
the loss flux of particles B at site $x$ at $t$. Noting that
B-particle that is at (or leaves)  site $x$ at time $t$ either has
come there as a B-particle at some prior time or was converted from
an A-particle that either was on site $x$ from the very beginning or
arrived there later at $t'>0$, and still keeps at (or just leaves)
the site $x$ at time $t$, we give the following balance
equations:
\begin{eqnarray} &B(x,t)&=
\int_{-\infty}^{+\infty}dx'\int_{0}^{t}j^{-}(x',t')\phi(x-x',t-t';x')dt'\nonumber\\
&&+\int_{-\infty}^{+\infty}dx'\int_{0}^{t}i^{-}(x',t')\phi(x-x',t-t';x')\nonumber\\
&&(1-e^{-\alpha(t-t')})dt'+A_{0}(x)\Psi(t)(1-e^{-\alpha t}),
\end{eqnarray}
and
\begin{eqnarray}
&j^{-}(x,t)&=
\int_{-\infty}^{+\infty}dx'\int_{0}^{t}j^{-}(x',t')\eta(x-x',t-t';x')dt'\nonumber\\
&&+\int_{-\infty}^{+\infty}dx'\int_{0}^{t}i^{-}(x',t')\eta(x-x',t-t';x')\nonumber\\
&&(1-e^{-\alpha(t-t')})dt'+
A_{0}(x)\psi(t)(1-e^{-\alpha t}),
\end{eqnarray}
where the initial condition $B(x,t)=0$ was used. Laplace
$x\rightarrow k$ and Fourier $t\rightarrow u$ transforming of
the two equations (25) and (26) yields:
\begin{eqnarray}
B(k,u)&=&\int j^{-}(k',u)\phi(k,u;k-k')dk'+\int
i^{-}(k',u)\nonumber\\
&&\times[\phi(k,u;k-k')-\phi(k,u+\alpha;k-k')]dk'+\nonumber\\
&&(\Psi(u)-\Psi(u+\alpha))A_{0}(k),
\end{eqnarray}
and
\begin{eqnarray}
j^{-}(k,u)&=&\int
j^{-}(k',u)\eta(k,u;k-k')dk'+\int
i^{-}(k',u)\nonumber\\
&&\times[\eta(k,u;k-k')-\eta(k,u+\alpha;k-k')]dk'\nonumber\\
&&+
(\psi(u)-\psi(u+\alpha))A_{0}(k).
\end{eqnarray}
Comparing (4), (7), (27) and (28), one has
\begin{equation}
\frac{A(k,u)+B(k,u)}{\Psi(u)}=\frac{i^{-}(k,u)+j^{-}(k,u)}{\psi(u)},
\end{equation}
which can be changed to the form:
\begin{equation}
j^{-}(k,u)=\Phi_{0}(u)B(k,u)+(\Phi_{0}(u)-\Phi_{\alpha}(u))A(k,u).
\end{equation}
 When $v(x)=0$ it is consistent with the result obtained in \cite{SSS2006}.

If the balance equation (25) is replaced by
\begin{eqnarray}
\frac{\partial B(x,t)}{\partial t}&=&\int_{-\infty}^{+\infty}j^{-}(x',t)\lambda(x-x'-\tau_{a}v(x'))dx'\nonumber\\
& &-j^{-}(x,t)+\alpha
A(x,t),
\end{eqnarray}
 we can also find Eq.(30)
 by using (7),(18),(28), the transform
$(x,t)\rightarrow (k,u)$ of Eq.(31)
\begin{eqnarray}
B(k,u)&=&\frac{\int
j^{-}(k',u)\eta(k,u+\alpha;k-k')dk'}{\psi(u+\alpha)}\nonumber\\
& &-j^{-}(k,u)+\alpha
A(k,u).
\end{eqnarray}
and the relation $\Psi(u)=\frac{1-\psi(u)}{u}$. It means that the two ways using different
 balance conditions for B-particles are equivalent, too.

In linear flow $v(x)=\omega x$ Eq.(32) can be written in the form:
\begin{equation}
B(k,u)\simeq\lambda(k)j^{-}(k+v_{k},u)-j^{-}(k,u)+\alpha A(k,u)
\end{equation} for small $\tau_a$.
Substitution of (29) into Eq.(33) gives the master equation for the
PDF of B-particle in Fourier-Laplace space:
\begin{eqnarray}
uB(k,u)=\lambda(k)v_{k}B'_{k}(k,u)\Phi_{0}(u)+\lambda(k)v_{k}A'_{k}(k,u)\nonumber\\
(\Phi_{0}(u)-\Phi_{\alpha}(u))+(\lambda(k)-1)B(k,u)\Phi_{0}(u)\nonumber\\
+(\lambda(k)-1)A(k,u)(\Phi_{0}(u)-\Phi_{\alpha}(u))+\alpha
A(k,u).
\end{eqnarray}
Inverting the above equation to the space-time domain, we obtain the
master equation in space-time domain:
\begin{widetext}
\begin{eqnarray}
&\frac{\partial B(x,t)}{\partial t}+\tau_{a}\int_{-\infty}^{+\infty}dx'\int_{0}^{t}\Phi_{0}(t-t')\lambda(x-x')\frac{\partial
v(x')B(x',t')}{\partial x'}dt'&\nonumber\\
&+\tau_{a}\int_{-\infty}^{+\infty}dx'\int_{0}^{t}[\Phi_{0}(t-t')-\Phi_{\alpha}(t-t')]\lambda(x-x')\frac{\partial
v(x')A(x',t')}{\partial x'}dt'=\int_{-\infty}^{+\infty}dx'\int_{0}^{t}[\Phi_{0}(t-t')-\Phi_{\alpha}(t-t')][\lambda(x-x')&\nonumber\\
&-\delta(x-x')]A(x',t')dt'+\int_{-\infty}^{+\infty}dx'\int_{0}^{t}\Phi_{0}(t-t')[\lambda(x-x'-\delta(x-x')]B(x',t')dt'+\alpha A(x,t).&
\end{eqnarray}
\end{widetext}
For a discrete random walk with $v(x)=0$, and the jump PDF satisfies
$\lambda(-1)=\frac{1}{2},\lambda(1)=\frac{1}{2}$,  in the continuum
limit, Eq.(35) reduces to Eq.(27) in Ref.~\cite{SSS2006}. If we substitute the
Gaussian jump length $ \lambda(k)\sim 1-\frac{\sigma^{2}k^{2}}{2}$
and
  power law waiting time PDF
$ \psi(u)\sim 1-\Gamma(1-\beta)(\tau u)^{\beta}$ into the master
equation (34), and invert $k\rightarrow x, u\rightarrow t$, in the
limit of small $\tau_{a}, \tau$ and $\sigma$, we then find the
generalized master equation for B particle in the reaction
$A\rightarrow B$ under subdiffusion in linear velocity field
\begin{eqnarray}
&&\frac{\partial B(x,t)}{\partial t}
+C_{\beta}[_{0}D_{t}^{1-\beta}-\hat{T}_{t}(1-\beta,\alpha)]\frac{\partial(v(x)A(x,t))}{\partial
x}\nonumber\\
&&+C_{\beta
}\cdot\,\leftidx_{0}D_{t}^{1-\beta}\frac{\partial(v(x)B(x,t))}{\partial x}\simeq K_{\beta
}\cdot\,\leftidx_{0}D_{t}^{1-\beta}\frac{\partial^{2}B(x,t)}{\partial x^{2}}\nonumber\\
&&+K_{\beta}[_{0}D_{t}^{1-\beta}-\hat{T}_{t}(1-\beta,\alpha)]\frac{\partial^{2}A(x,t)}{\partial
x^{2}}+\alpha A(x,t).
\end{eqnarray}
It should be noted that in (36) both  advection and diffusion terms
for B-particle couple with the reaction too, since these two terms
respectively  include not only the advection term
$C_{\beta}\hat{T}_{t}(1-\beta,\alpha)\frac{\partial(v(x)A(x,t))}{\partial
x}$ and diffusion term
$K_{\beta}\hat{T}_{t}(1-\beta,\alpha)\frac{\partial^{2}A(x,t)}{\partial
x^{2}}$ in (23), but also the fractional kinetics memories  at
previous times for A-particle.

  Let
$C(x,t)$ denote the sum of $A(x,t)$ and $B(x,t)$. Combining (23)
with (36), one has
\begin{eqnarray}
&\frac{\partial C(x,t)}{\partial t}+C_{\beta
}\cdot\,\leftidx_{0}D_{t}^{1-\beta}\frac{\partial(v(x)C(x,t))}{\partial x} &\nonumber\\
&= K_{\beta
}\cdot\,\leftidx_{0}D_{t}^{1-\beta}\frac{\partial^{2}C(x,t)}{\partial x^{2}}&
\end{eqnarray}
which is consistent with the result for subdiffusion in nonreactive liquid in \cite{C1997,CMC1997}.
It is because that the simple reaction we discuss here does not
change the sum of the particles in the system.

To sum up we derive the master equations (20) and (35) for the PDF
of A and B particles in a simple monomolecular conversion reaction
$A\rightarrow B$ taking place at a constant rate $\alpha$ and under
subdiffusion in linear flows. As examples, two generalized
advection-diffusion reaction equations (23) and (36) are obtained,
and the interesting couple relations among diffusion, advection and
reaction process are showed.  There are problems such as the dynamic
behaviors for more complex reaction under subdiffusion in moving
fluids are still unknown.

\begin{acknowledgements}
This work was supported by the National Natural Science Foundation of China (Grant
No. 11626047) and the Foundation for Young Key Teachers of Chengdu
University of Technology, China (Grant No. KYGG201414).
\end{acknowledgements}

\bibliographystyle{apsrev}
\bibliography{我的}

\end{document}